\documentclass[twocolumn,aps,superscriptaddress,showpacs,nofootinbib,floatfix]{revtex4}
\usepackage{epsfig,bm,feynmf}
\usepackage{graphics}
\usepackage{amsmath}
\usepackage[normalem]{ulem}  % \sout{old text} for strikeout
\usepackage[dvips]{color} % For blue in-text comments and additions

\renewcommand\sout{\bgroup \color{red} \ULdepth=-.5ex \ULset}
\begin{document}

%%%%%%%%%%%%%%%%%%%%% Title %%%%%%%%%%%%%%%%%%%%%%

\title{Quarkonium formation time in quark-gluon plasma}

%%%%%%%%%%%%%%%%%%%% Authors %%%%%%%%%%%%%%%%%%%%%

\author{Taesoo Song}\email{tsong@comp.tamu.edu}
\affiliation{Cyclotron Institute, Texas A$\&$M University, College Station, TX 77843, USA}
\author{Che Ming Ko}\email{ko@comp.tamu.edu}
\affiliation{Cyclotron Institute and Department of Physics and Astronomy, Texas A$\&$M University, College Station, TX 77843, USA}
\author{Su Houng Lee}\email{suhoung@yonsei.ac.kr}
\affiliation{Institute of Physics and Applied Physics, Yonsei
University, Seoul 120-749, Korea}

%%%%%%%%%%%%%%%%%%%% Abstract %%%%%%%%%%%%%%%%%%%%%

\begin{abstract}
The quarkonium formation time in a quark-gluon plasma (QGP) is determined from the space-time correlator of heavy quark vector currents using the quarkonium in-medium mass and wave function obtained from heavy quark potentials extracted from the lattice QCD. It is found that the formation time of a quarkonium increases with the temperature of the QGP and diverges near its dissociation temperature. Also, the quarkonium formation time is longer if the heavy quark potential is taken to be the free energy from lattice calculations for a heavy quark pair, compared to that based on the more negative internal energy.
\end{abstract}


\maketitle

\section{introduction}

Quarkonia of heavy quark-antiquark bound states are promising probes for the QGP created in relativistic heavy-ion collisions~\cite{Matsui:1986dk}. Many studies have been carried out in recent years. These include experimental measurements of its yield in heavy ion collisions at both the Relativistic Heavy Ion Collider (RHIC)~\cite{Adare:2006ns,Adare:2008sh,Abelev:2009qaa,Adare:2011yf} and the Large Hadron Collider (LHC)~\cite{:2010px,Chatrchyan:2011pe,Abelev:2012rv}, as well as theoretical studies based on various models~\cite{Vogt:1999cu,Zhang:2000nc,Zhang:2002ug,Grandchamp:2002wp,Yan:2006ve,Zhao:2007hh,Song:2010er,Song:2011xi,Song:2012at,Song:2011nu,Song:2013tla}. These studies have indicated that a quantitative description of quarkonium production in these collisions requires the understanding of its interactions in both the produced QGP and the initial cold nuclear matter. One of the important QGP effects is the dissociation of quarkonium by thermal patrons~\cite{Park:2007zza,Song:2007gm,Zhao:2007hh}. The resulting thermal dissociation width of a quarkonium is determined by its dissociation cross section and the density of partons. In both the leading order~\cite{Peskin:1979va,Bhanot:1979vb,Oh:2001rm} and the next-to-leading order calculations~\cite{Song:2005yd} in perturbative Quantum Chromodynamics (pQCD), the quarkonium dissociation cross section is characterized by its dipole nature through the derivative of the quarkonium wave function with respect to the relative momentum between the heavy quark and antiquark. However, the thermal dissociation of quarkonium in QGP is relevant only after the quarkonium is completely formed from the heavy quark-antiquark pair that are produced from initial hard scatterings between colliding nucleons. While the heavy quark pair are produced during the time of $1/m_Q$ with $m_Q$ being the heavy quark mass, a quarkonium takes a longer time to be formed from the pair. This is because the typical time scale needed to separate the bound state from the continuum or the excited states are related to the inverse of their energy differences, which in the heavy quark limit of QCD scales as the bound state energy $m_Q g^4$, where $g$ is the QCD coupling constant~\cite{Oh:2001rm}.

There have been several attempts to determine the formation time of quarkonium~\cite{Blaizot:1988ec,Karsch:1987zw,Kharzeev:1999bh}. One of them is based on the space-time correlator of the heavy quark current operator~\cite{Kharzeev:1999bh}. This approach has the advantage that all parameters involved in the calculation can be related to experimentally measured quantities in heavy quark production from electron-positron annihilation, $e^+ e^- \rightarrow \bar{Q}Q$. In the present study, we extend this method to study the quarkonium formation time in QGP by using the quarkonium in-medium mass and wave function that are determined by taking the heavy quark potential to be either the free energy or the internal energy of a heavy quark pair in QGP as extracted from lattice calculations~\cite{Digal:2005ht,Satz:2005hx,Kaczmarek:1900zz}.

This paper is organized as follows: In Sec.~\ref{formulas}, we briefly review the quarkonium formation time in the approach based on the space-time correlator of the heavy quark vector current. We then describe in Sec.~\ref{results} the quarkonium formation time in QGP by using the quarkonium in-medium mass and wave function calculated from the lattice heavy quark potentials. Finally a summary is given in Sec.~\ref{summary}.

\section{formation time of quarkonum}\label{formulas}

For the heavy quark vector current operator $J_\mu(x)={\bar Q}\gamma_\mu Q$, its space-time correlator $\Pi_{\mu\nu}(x)$ is given by
\begin{eqnarray}
\Pi_{\mu\nu}(x)&=&\langle 0| T\{J_\mu(x)J_\nu(0)\}|0\rangle\nonumber\\
&=&\int \frac{d^4q}{(2\pi)^4}e^{-iq\cdot x}(q_\mu q_\nu-g_{\mu\nu}q^2)\Pi(q^2).
\label{correlator}
\end{eqnarray}
Using the  dispersion relation
\begin{eqnarray}
\Pi(q^2)=\frac{1}{\pi}\int ds\frac{{\rm Im}\Pi(s)}{s-q^2}
\label{dispersion}
\end{eqnarray}
to relate the real part of the heavy quark correlator in energy-momentum space or pair polarization function $\Pi(q^2)$ to its imaginary part, one can rewrite, for $x\ne 0$, Eq.~(\ref{correlator}) as~\cite{Shuryak:1983jg,Shuryak:1986uu,Kharzeev:1999bh}
\begin{eqnarray}
\Pi(x)\equiv\Pi_\mu^\mu(x)&=&\frac{3}{\pi}\int dss~ {\rm Im}\Pi(s)\int \frac{d^4q}{(2\pi)^4}\frac{e^{-iq\cdot x}}{q^2-s}\nonumber\\
&=&\frac{3}{\pi}\int dss ~{\rm Im}\Pi(s)D(s,x^2),
\label{correlator2}
\end{eqnarray}
where
\begin{eqnarray}
D(s,\tau^2=-x^2)=\frac{\sqrt{s}}{4\pi^2\tau}K_1(\sqrt{s}\tau)
\end{eqnarray}
is the relativistic causal propagator in the coordinate space with $K_1$ being the modified Bessel function and $\tau$ being the Euclidean proper time~\cite{Bogoliubov}.

The imaginary part of the heavy quark pair polarization function is related to the ratio of the cross section for heavy quark pair production to that for dimuon production in electron-positron annihilation,
\begin{eqnarray}
{\rm Im}\Pi(s)=\frac{1}{12\pi}\frac{\sigma(e^+e^-\rightarrow Q\bar{Q},s)}{\sigma(e^+e^-\rightarrow \mu^+\mu^-,s)}.
\label{imaginary}
\end{eqnarray}
The cross section ratio $R(s)\equiv\frac{\sigma(e^+e^-\rightarrow Q\bar{Q},s)}{\sigma(e^+e^-\rightarrow \mu^+\mu^-,s)}$ in the above equation can be parameterized by the sum of resonances and a continuum~\cite{Kharzeev:1999bh}:
\begin{eqnarray}
R(s)&=&\sum_i (2J_i+1)\frac{3}{4\alpha^2}\frac{\Gamma_i^{e^+e^-}\Gamma_i}{(\sqrt{s}-M_i)^2+(\Gamma_i/2)^2}\nonumber\\
&&+3e_Q^2\theta(s-s_{\rm th}),
\label{R-ratio}
\end{eqnarray}
where $J_i$, $M_i$, $\Gamma_i^{e^+e^-}$, and $\Gamma_i$ are the spin, mass, dielectron and total decay widths of resonance $i$, respectively; $e_Q$ is the electric charge of the heavy quark and the threshold energy $\sqrt{s_{\rm th}}$ is taken to be twice the mass of open heavy flavor.

Substituting Eqs.~(\ref{imaginary}) and (\ref{R-ratio}) into Eq.~(\ref{correlator2}), one obtains following contributions to the heavy quark space-time correlator from resonances and the continuum~\cite{Kharzeev:1999bh}:
\begin{eqnarray}
\Pi_{\rm res}(\tau)&=&\sum_i\frac{3(2J_i+1)\Gamma_i^{e^+e^-}M_i^4}{16\pi^3\tau\alpha^2}K_1(M_i\tau),\nonumber\\
\Pi_{\rm cont}(\tau)&=&\frac{3e^2_Q}{8\pi^4\tau^6}\int^\infty_{\sqrt{s_{\rm th}}\tau}x^4K_1(x)dx.
\label{pi}
\end{eqnarray}
These expressions show that the contribution from the continuum dominates at earlier times and is then gradually taken over by that from resonances. This is consistent with the picture that the $Q\bar{Q}$ pair produced in $e^+ e^-$ annihilation is initially formed in a continuum state and then evolves into a quarkonium state.

By taking the ratio of $F_i(\tau)=\Pi_i(\tau)/\Pi_>(\tau)$, where $\Pi_i(\tau)$ is the heavy quark space-time correlator due to resonance $i$ and $\Pi_>(\tau)$ is the total heavy quark pair polarization function excluding contributions from quarkonium resonances with masses lower than the mass of quarkonian $i$, one can then obtain the distribution of the formation time of quarkonium $i$ according to~\cite{Kharzeev:1999bh}
\begin{eqnarray}
P_i(\tau)=\frac{dF_i(\tau)}{d\tau}.
\end{eqnarray}

\section{quarkonium formation time in QGP}\label{results}

Although the quantities $M_i$, $\Gamma_i^{e^+e^-}$, and $\sqrt{s_{\rm th}}$ appearing in Eq.~(\ref{pi}) are well known if one is interested in determining the formation time of a quarkonium in free space, their values in hot dense medium depends on theoretical models. In the present study, we determine them using the heavy quark potential taken from the potential energy between a heavy quark-antiquark pair that is extracted from lattice calculations~\cite{Digal:2005ht,Satz:2005hx,Kaczmarek:1900zz}. While the heavy quark potential based on their free energy is directly extracted from lattice calculations, that based on their internal energy depends also on the entropy density. As a result, the internal energy potential is stronger than the free energy potential. At present, it is still an open question as to which potential energy should be used in determining the properties of a quarkonium in QGP. We therefore use both potential energies for the present study and compare the resulting predictions.

For a given potential between a heavy quark and antiquark pair, the Schr\"odinger equation has the form
\begin{eqnarray}
\bigg[2m_Q-\frac{1}{m_Q}\nabla^2+V(r,T)\bigg]\psi_i(r,T)=M_i\psi_i(r,T),
\end{eqnarray}
where $m_Q$ is the bare mass of heavy quark $Q$ and $\psi_i(r,T)$ is the wave function of quarkonium $i$ at temperature $T$.
Introducing the potential energy at infinity, $V(r=\infty, T)$, the Schr\"odinger equation is modified to~\cite{Karsch:1987pv,Satz:2005hx}
\begin{eqnarray}
&&\bigg[-\frac{1}{m_Q}\nabla^2+\widetilde{V}(r,T)\bigg]\psi_i(r,T)\nonumber\\
&&~~~=-\bigg\{2m_Q+V(r=\infty, T)-M_i\bigg\}\psi_i(r,T)\nonumber\\
&&~~~=-\bigg\{2\widetilde{m}_Q-M_i\bigg\}\psi_i(r,T)\nonumber\\
&&~~~=-\varepsilon_i\psi_i(r,T),
\label{schrodinger2}
\end{eqnarray}
where $\widetilde{V}(r,T)\equiv V(r,T)-V(r=\infty, T)$ and it vanishes at infinity, $\widetilde{m}_Q\equiv m_Q+V(r=\infty, T)/2$, and $\varepsilon_i$ is the binding energy of quarkonium $i$ at temperature $T$.

From its wave function at the origin, the dielectron decay width of quarkonium $i$ can be calculated according to~\cite{Oh:2001rm}
\begin{eqnarray}
\Gamma_i^{e^+e^-}=\frac{16\pi\alpha^2 e_Q^2}{M_i^2}~|\psi(r=0)|^2,
\end{eqnarray}
where $\alpha$ is the fine structure constant.

\begin{figure}[h]
\centerline{
\includegraphics[width=9 cm]{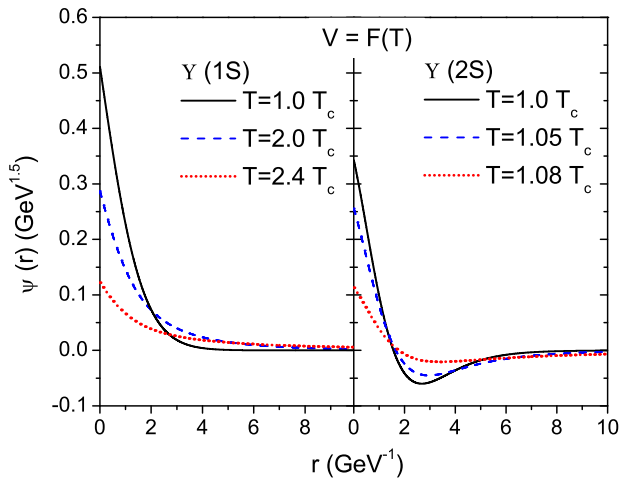}}
\caption{(Color online) Wave functions of $\Upsilon$ (1S) and $\Upsilon$ (2S) for the case of free energy potential at $T=1.0,~2.0,~2.4~T_c$ and at $T=1.0,~1.05,~1.08~T_c$, respectively.}
\label{wavefts}
\end{figure}

As an example, we consider the formation time of bottomonia in QGP. Their in-medium masses and wave functions are obtained by solving Eq.~(\ref{schrodinger2}) with the bottom quark mass of $m_b=$ 4.65 GeV. Figure~\ref{wavefts} shows their wave functions of $\Upsilon(1S)$ and $\Upsilon(2S)$ for the free energy potential at several temperatures. It is seen that as temperature increases, the wave function of quarkonium becomes broader. As a result, the probability for the heavy quark and antiquark in the quarkonium to annihilate and decay into dielectron becomes smaller. We note that the binding energies of $\Upsilon(1S)$ and $\Upsilon(2S)$ are larger and their wave functions also have larger values at the origin if the more attractive heavy quark potential based on the internal energy is used.

\begin{figure}[h]
\centerline{
\includegraphics[width=9 cm]{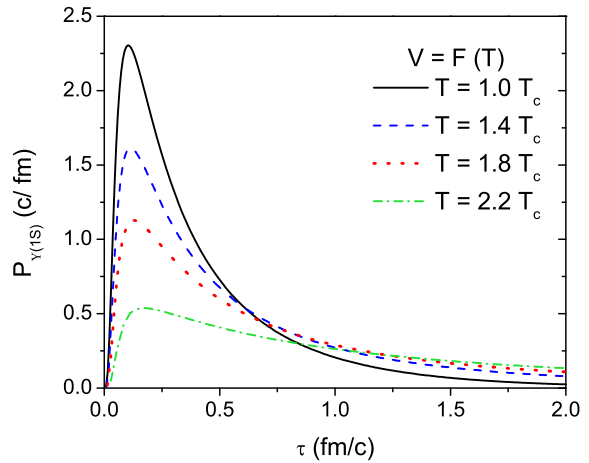}}
\caption{(Color online) Distribution of $\Upsilon$ (1S) formation time for the case of free energy potential at $T=1.0,~1.4,~1.8$, and $2.2~T_c$.}
\label{distributions}
\end{figure}

Taking the threshold energy for the continuum contribution in Eq.(\ref{pi}) to be $\sqrt{s_{\rm th}}=2\widetilde{m}_Q$, we have calculated the distribution of the $\Upsilon$ (1S) formation time at several temperatures for the case of free energy potential, and they are shown in Fig.~\ref{distributions}. The distribution peaks at small $\tau$ at $T=1.0~T_c$ but spreads to later times at high temperature. There are two reasons for this behavior of the $\Upsilon$ formation time in QGP. First, the smaller $\Gamma_i^{e^+e^-}$ at high temperature delays the time for the contribution of quarkonium to the heavy quark correlator $\Pi(\tau)$ to become dominant. Secondly, the small binding energy of quarkonium at high temperature makes its formation to take longer time. In the large Euclidean time $\tau$ limit, the contributions to the heavy quark correlator from a resonance $i$ and the continuum are proportional to $e^{-M_i\tau}$ and $e^{-\sqrt{s_{\rm th}}\tau}$, respectively~\cite{Kharzeev:1999bh}. Because the resonance mass $M_i$ is always smaller than the threshold for the continuum, the contribution from a resonance eventually becomes dominant. However, if the binding energy of quarkonium is very small, it takes a long time for the contribution of the quarkonium state to become much more important than that of the continuum states.

\begin{figure}[h]
\centerline{
\includegraphics[width=9 cm]{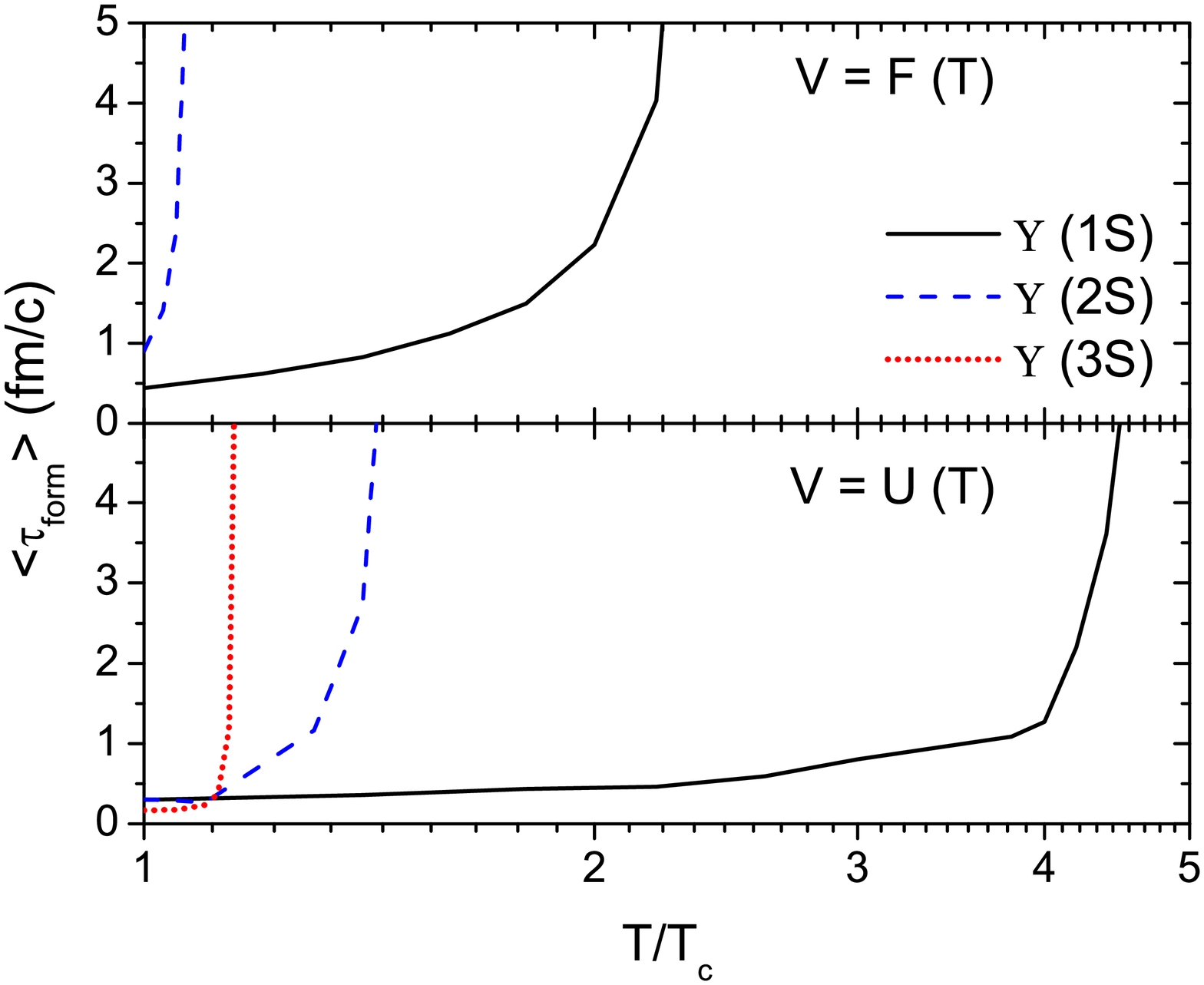}}
\caption{(Color online) Average formation times of $\Upsilon(1S)$, $\Upsilon(2S)$, and $\Upsilon(3S)$ as functions of temperature for the cases of free energy (upper) and internal energy (lower) potentials.}
\label{formations}
\end{figure}

Figure~\ref{formations} shows the average formation times of $\Upsilon(1S)$, $\Upsilon(2S)$, and $\Upsilon(3S)$ as functions of temperature,
\begin{eqnarray}
\langle\tau_{\rm form}\rangle=\frac{\int d\tau ~\tau P_i(\tau)}{\int d\tau P_i(\tau)},
\end{eqnarray}
for the two cases of free energy and internal energy potentials. It is seen that the average formation time of quarkonium increases with temperature and diverges near its dissociation temperature. This behavior is similar to that of the quarkonium radius in QGP~\cite{Song:2011nu}. This is reasonable because as the quarkonium radius increases at high temperature, it takes longer for the quarkonium to be formed~\cite{Karsch:1987zw}. Since the quarkonium is less bound and has a larger radius in the case of the free energy potential than that of the internal energy potential, the formation time is thus longer in the former than in the latter case. Compared to the formation time of 0.32 fm/$c$ obtained in Ref.~\cite{Kharzeev:1999bh} for a $\Upsilon(1S)$ in free space, our values of 0.44 fm/$c$ and 0.3 fm/$c$ at $T=T_c$ for the cases of free energy and internal energy potentials are slightly larger and similar, respectively.

\begin{figure}[h]
\centerline{
\includegraphics[width=9 cm]{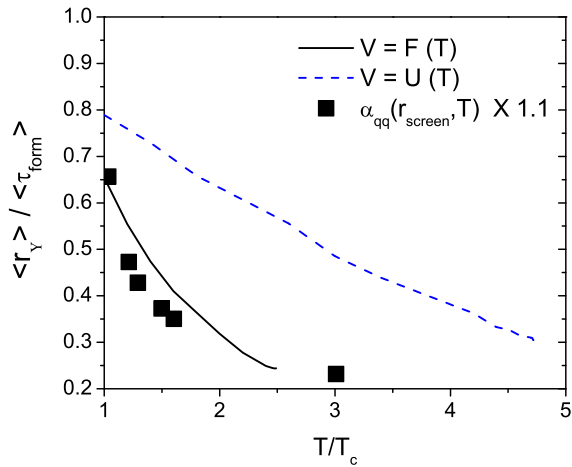}}
\caption{(Color online) Ratio of the mean distance between bottom quark and antibottom quark in $\Upsilon$ (1S) to its formation time as a function of temperature for the cases of free energy (solid line) and internal energy (dashed line) potentials. Solid squares are the temperature dependence of the QCD coupling constant (scaled by 1.1) at the screening distance between heavy quark and antiquark pair extracted from the lattice free energy.}
\label{velocities}
\end{figure}

In Fig.~\ref{velocities}, we show the ratio of the mean distance between bottom quark and antibottom quark in $\Upsilon$ (1S) to its formation time as a function of temperature, calculated according to $\int d^3{\bf x} r|\psi(r)|^2/\langle\tau_{\rm form}\rangle$.
It can be interpreted as the average relative velocity between bottom and antibottom quarks before they form the $\Upsilon$ (1S) state. Results in Fig.~\ref{velocities} show that the relative velocity is smaller for the case of free energy potential than for that of internal energy potential, and that it decreases as temperature increases. The latter is consistent with the fact that the wave function of bottomonium in momentum space has a small relative momentum at high temperature~\cite{Song:2010ix}. However, this does not necessarily mean that only bottom quark pairs with small relative momentum contribute to bottomonium formation at high temperature, because even if a heavy quark pair initially have a large relative momentum, they can lose energy during propagation through QGP~\cite{Adare:2010de,delValle:2011ex} and thus become slowly moving relative to each other.

It is useful to discuss the result shown in Fig.~\ref{velocities} using the relevant scales for the heavy quark system in QCD. The heavy quark system is characterized by three scales; hard $m_Q$, soft $p=m_Q v$ and ultra-soft $E=m_Q v^2$ in terms of the heavy quark mass and velocity. The different scales are the basis for the potential non-relativistic QCD (pNRQCD) approach, which is an effective field theory for heavy quarks with bound states~\cite{Brambilla:1999xf}. In the $m_Q \rightarrow \infty$ limit, one notes that the soft and ultra-soft scales are proportional to the inverse Bohr radius $1/a_0$ and the binding energy of the system, which scale with the QCD coupling as $m_Qg^2$ and $m_Qg^4$, respectively~\cite{Peskin:1979va}. Therefore, the heavy quark pair mean distance to the quarkonium formation time ratio shown in Fig.~\ref{velocities} should be proportional to $g^2$ or $\alpha_{\rm eff}(T)$. In Ref. \cite{Kaczmarek:2004gv}, $\alpha_{qq}(r_{\rm screen},T)$ was extracted from the lattice free energy by estimating the nonperturbative coupling constant at the screening distance between the quark and antiquark pair, which should be the most relevant distance for the heavy bound state. It is quite interesting to see that the temperature dependence of $\alpha_{qq}(r_{\rm screen},T)$, scaled by an overall factor of 1.1, is close to the ratio in Fig.~\ref{velocities} extracted from the free energy potential.

\section{summary}\label{summary}

We have studied the quarkonium formation time in QGP by using the approach based on the space-time correlator of heavy quark vector currents. The imaginary part of the resulting heavy quark pair polarization function, which is the spectral function of heavy quark pair in $e^+e^-$ annihilation, is constructed by solving the Schr\"odinger equation with the heavy quark potential extracted from lattice calculations. The real part of the polarization function, which is related to its imaginary part by the dispersion relation, then provides the information on how different states of the heavy quark pair evolve with time.

Using bottomonia as examples, we have found that the average formation time of a quarkonium from a heavy quark-antiquark pair increases with temperature and diverges near the dissociation temperature of the quarkonium. Furthermore, the quarkonium formation time is longer if the heavy quark potential is taken to be their free energy instead of their internal energy from the lattice calculations. We have also found that the average relative velocity between the heavy quark pair before they form the quarkonium, calculated via the ratio of the mean distance between the heavy quark and antiquark in a quarkonium to its formation time, decreases with increasing temperature. Our results thus indicate that because of the longer formation time of quarkonium at high temperature, to fully describe quarkonium production in relativistic heavy ion collisions requires a proper treatment of the medium effects on heavy quark pairs in the QGP before they form the quarkonium bound state.

\section*{Acknowledgements}

This work was supported in part by the U.S. National Science Foundation under Grant No. PHY-1068572, the Welch Foundation under Grant No. A-1358, and the Korean Research Foundation under Grant Nos. KRF-2011-0030621 and KRF-2011-0015467.

\end{document}